\documentclass[twocolumn,showpacs,aps,prc,superscriptaddress]{revtex4-1}

\usepackage{amsmath, bm}
\usepackage{epsfig}
\usepackage{color}
\usepackage{soul}
\newcommand{\sNN}{$\sqrt{s_{_\mathrm{NN}}}$ }
\newcommand{\mupi}{$\mu$-$\pi$ }
\newcommand{\muk}{$\mu$-$K$ }
\newcommand{\mup}{$\mu$-$p$ }

\newcommand{\KL}{$K^{0}_{L}~$}

\begin{document}

\title{Coalescence formation of muonic atoms at RHIC}

\author{Xiaofeng Wang}
\email{xiaofeng\_wang@mail.sdu.edu.cn}
\affiliation{Key Laboratory of Particle Physics and Particle Irradiation (MOE), Institute of Frontier and Interdisciplinary Science, Shandong University, Qingdao, China 266237}

\author{Frank Geurts}
\email{geurts@rice.edu}
\affiliation{T.W. Bonner Nuclear Laboratory, Rice University, Houston, TX77005-1892, USA}

\author{Zebo Tang}
\email{zbtang@ustc.edu.cn}
\affiliation{Department of Modern Physics, University of Science and Technology of China, Hefei
230026, China}

\author{Kefeng Xin}
\email{allencopy@gmail.com}
\affiliation{T.W. Bonner Nuclear Laboratory, Rice University, Houston, TX77005-1892, USA}

\author{Zhangbu Xu}
\email{zxu22@kent.edu}
\affiliation{Physics Department, Kent State University, Kent, OH 44242, USA}
\affiliation{Physics Department, Brookhaven National Laboratory, Upton, NY11973, USA}

\author{Yifei Zhang}
\email{ephy@ustc.edu.cn}
\affiliation{Department of Modern Physics, University of Science and Technology of China, Hefei
230026, China}

\author{Long Zhou}
\email{zhoulong07@hotmail.com}
\affiliation{Department of Modern Physics, University of Science and Technology of China, Hefei
230026, China}


\begin{abstract}
The discovery of exotic mounic atoms, including antimatter hydrogen muonic atoms and kaon mounic atoms, constitutes a milestone in our ability to make and study new forms of matter. Relativistic heavy-ion collisions provide the only likely condition for production and detection of these exotic atoms. Taking a Coulomb correlations into account from the time of the fireball freeze-out until the formation of a stable atom has dramatic consequence on the expected yields of these atoms. When the coalescence model with the assumption of quantum wave function localization is applied to the formation of muonic atoms, we find that the atom yields are about two orders of magnitude higher than previously predicted.

\end{abstract}

\date{\today}
 \maketitle

\maketitle

Muonic atoms with muons captured by protons or nuclei have been studied over many decades
and have become a precision tool in fundamental physics measurements that involve, \textit{e.g.}, the
determination of nuclear shapes and masses \cite{Antognini:2013txn}. However, muonic atoms with the nuclear
core replaced by more exotic particles (pions, kaons, or antiprotons) are rarely produced.
These hydrogen-like muonic atoms are $\mu$-$\pi$,  \muk and \mup  (antimatter muonic hydrogen).
To date, only one such an object, the \mupi system, has been observed: at Brookhaven National Lab in 1976 by R.~Coombes
{\it et al.} \cite{Coombes:1976hi}(33 events), and at Fermilab in 1982 by S.H.~Aronson {\it et al.} \cite{Aronson:1982bz} (320 events).
In these experiments, the atoms were formed from \KL\ decays. A high-intensity kaon beam, in which the decay
products muon and pion are formed at low relative momentum, provides the necessary conditions for formation and observation.
The apparent production rate of $4\times10^{-7}$ per \KL\ is consistent with a prediction from the wave-function overlap at zero separation computed using only the Coulomb interaction. We propose a more direct approach to the production of new kinds of exotics atoms which takes advantage of the large number of thermal muons and hadrons produced in a relativistic heavy-ion collision, followed by the detection of these particles in the STAR (Solenoidal Tracker At RHIC) detector at the Relativistic Heavy Ion Collider (RHIC), using its relative yields to determine the conditions in the early stages of the collision.

Mel Schwartz was the first to propose measuring the distributions of exotic atoms formed by binding a directly produced lepton to one of the charged hadrons emerging from the final state of a nuclear collision. The idea is not just to discover the exotic atoms, but also to measure the thermal electromagnetic emission from the quark-gluon plasma (QGP) via a direct measurement of the single muon spectrum.  Measurements in this range are capable of probing thermal electromagnetic emission processes in an initial plasma with a temperature of a couple of hundred MeV, as argued by Gordon~Baym and others~\cite{Baym:1993ae}. The soft leptons produced in the early evolution of Au$+$Au collisions at RHIC are difficult to measure because of the large number of charged particles created in the collisions, which eventually weakly decay to muons.
Muons produced from most of the weak hadronic decays are not captured by the
produced hadrons to form atoms because the decays happen at a very late stage. The exotic atoms are produced by hadrons
and thermal muons or muons from resonance decay (\textit{e.g.}, $\rho\rightarrow\mu^+\mu^-$), which exist right around freeze-out.
In ultra-relativistic heavy-ion collisions, strong electromagnetic fields arising from the Lorentz-contracted, highly charged nuclei can be approximated as a large flux of high-energy quasi-real photons that can interact via the Breit-Wheeler process to produce $\mu^{+}\mu^{-}$ pairs~\cite{vonWeizsacker:1934nji,Williams:1934ad,ATLAS:2022yad,ATLAS:2018pfw,STAR:2018ldd,STAR:2019wlg,STAR:2024qpx,Lin:2022flv}
The muons produced from the Breit-Wheeler process in the early stages of the collision may also fuse with hadrons to form exotic atoms.
Muons from weak or electromagnetic decays are produced too far away from the collision zone to be combined with the collision
hadrons. The exception is the case of $K_L^0$ decay. However, the $K_L^0$  yields and the fraction of its decay within the STAR
detector make negligible contribution to the total \mupi yields. In addition, production from other atoms (\muk and $\mu$-$p$)
will provide a unique signal for the formation of such atoms from the collisions.
Detailed rate estimates were carried out by Baym \textit{et al.} \cite{Baym:1993ae}, and by Kapusta and
M\'ocsy~\cite{Kapusta:1998fh}, showing sensitivity to initial momentum distributions and particle densities. Since the collision zone at freeze-out is on the order of 10~fm in dimension while the size of the formed atoms is of hundreds of fm, only the wave functions of the atom's $S$ states at the origin contribute to the formation probability~\cite{Baym:1993ae}.

\begin{equation}
\frac{dN_{\rm atom}}{{dyd^2p_{T, \mathrm{atom} } }}= 8\pi^2\zeta (3) \alpha^3 m^2_{\rm red} \frac{dN_h}{dyd^2p_{T,h}} \frac{dNl}{dyd^2p_{T,l}}
\label{ex:equation:baym}
\end{equation}

Equation~\eqref{ex:equation:baym} shows that the yield of atoms is directly proportional to the yields of hadrons and leptons in the collision event. The factor is dominated by the QED fine structure constant $\alpha^3 \simeq 4\times10^{-7}$. The copiously produced hadrons and muons in relativistic heavy-ion collisions make it a unique system to produce exotic atoms such as antimatter muonic hydrogen. Baym \textit{et al.}~\cite{Baym:1993ae} have provided an estimate for the integrated yield of $\sim10^{-5}$ \mupi atoms in one unit of rapidity per central Au+Au collision. This means that STAR samples about 10,000 such events when saving 500 million central Au+Au collision events to tape. Kapusta and M\'ocsy~\cite{Kapusta:1998fh} extended the calculation to $p_T$ spectra, and the estimate is about $10^{-5}$ \mupi  atoms at $p_T \simeq 1 \pm 0.5$~GeV/$c$ per central Au+Au collisions. In both calculations, the hadrons and muons are assumed to be from a thermal distribution, and the muon yield is estimated to be a factor of 5000 lower than the pion yield.

However, previous studies did not take into account the Coulomb potential at the freeze-out in relativistic heavy-ion collisions. In \cite{Baym:1993ae},
it was stipulated that in a hydrogenic atom, the relative velocity of the two constituents is $\alpha c/n$, the Bohr velocity, where $n$ is the principal quantum number, and that two oppositely charged constituents in the collision volume will form an atom only if their relative velocity at freeze-out is $\lesssim\alpha c$.
This effectively assumes that the particles have no correlations because their wave functions were not localized at the time of the creation.
However, results from relativistic heavy-ion experiments have shown, ({\it e.g.}, the recent study of antiproton interactions~\cite{STAR:2015kha} and pion-kaon correlations~\cite{STAR:2003cqe}) that the Coulomb interaction at freeze-out greatly enhances the correlation function of oppositely charged particles. The theoretical framework has been laid out two decades ago by Baym and Braun-Munzinger~\cite{BAYM1996286}. In their studies, a toy model with the assumption of a classic Coulomb interaction at freeze-out is able to explain the experimental data of charged pion correlations at small relative momenta.
The correlation function $C({\vec{q})}$ at infinity distance is written as:
\begin{equation}
C(\vec{q})=\sqrt{1\pm\frac{2m_{\rm red}e^{2}}{r_{0}q^{2}}} \; C_{0}(\vec{q}_{0}).
\label{coulomb}
\end{equation}
where $\vec{q}$ is the relative momentum at infinite distance, $\vec{q}_{0}$ and $r_{0}$ are
the relative momentum and distance of the two particles at freeze-out when hadronic interactions within the fireball cease.
For pairs with initial distances larger than the classic turning point, its propagation as free particles at infinity and the connection to the wave-function description in quantum mechanics have also been discussed in the literature~\cite{BAYM1996286}.
Although these two papers~\cite{Baym:1993ae,BAYM1996286} have discussed two different cases of interactions, {\it i.e.}
the formation of a bound state of a pair by Coulomb potential and the correlation function of two charged particles,
the treatment of the phase-space distribution of the pairs at atomic scale is very different.
In Ref.~\cite{Baym:1993ae}, the wave functions are assumed to be scattering waves, and independent of each other regardless of where the particles are created.
In Ref.~\cite{BAYM1996286}, however, it is assumed that the quantum-wave functions of all the particles are localized at the fireball freeze-out.
The latter treatment~\cite{BAYM1996286} results in a dramatic enhancement of the pair of opposite charges in the phase space at atomic scale due to strong Coulomb correlations at the initial creation at distance ($r_{0}$).

Since the classic picture in this toy model~\cite{BAYM1996286} describes the experimental correlation data as a function of the relative momentum of the two consitituents very well for an unbound state, we apply a similar approach to the bound state in the coalescence model. The coalescence model has worked well in describing the formation of nuclear clusters in relativistic heavy-ion collisions~\cite{STAR:2010gyg,STAR:2011eej,E864:1999zqh}. In the simple coalescence of two nucleons into a deuteron~\cite{Butler:1961pr,Nagle:1996vp,Llope:1995zz}, the formula is as follows:
\begin{equation}
\frac{dN_{d}}{{dyd^2p_{T , d} }}= B_2\frac{dN_p}{dyd^2p_{T ,p}}\frac{dN_n}{dyd^2p_{T ,n}}
\label{deuteron}
\end{equation}
where
\begin{equation}
B_2= \frac{3}{4}\,\Big( \frac{4\pi}{3}p_{0}^{3} \Big) \, \frac{1}{m_{red}},
\label{B2}
\end{equation}
similarly, for muonic atoms,
\begin{equation}
B_2= \Big( \frac{4\pi}{3}p_{0}^{3} \Big) \, \frac{1}{m_{red}},
\label{B2}
\end{equation}
${dN_{d}}/{{dyd^2p_{T , x} }}$ denote the spectra of the nuclei and its constituents respectively, 
and $p_0$ is the maximum relative momentum cut-off for the two particles to coalesce.

In this framework, we assume that the muonic atom will form if the relative momentum of the muon and the hadron at the fireball freeze-out is not enough to overcome their Coulomb potential. That is:
\begin{equation}
\frac{q^2_{0}}{2m_{\rm red}}\leq \frac{1}{4\pi\varepsilon_{0}} \, \frac{e^2}{r_{0}},
\label{space}
\end{equation}
where $\varepsilon_{0}$ is vacuum permittivity.

Given that the relative momentum of the constituents of the muonic atom is much smaller than the initial relative momentum $q_{0}$, this allows for a much larger phase space for a muon and hadron to coalesce into a muonic atom.
Taking this phase space into account in the coalescence model~\cite{Butler:1961pr,Llope:1995zz,Nagle:1996vp,Schwarzschild:1963zz} at freeze-out and neglect other momentum and space correlations (such as, flow and minijets),
we arrive at:
\begin{equation}
\frac{dN_{\rm atom}}{{dyd^2p_{T , \mathrm{atom} } }}= \frac{4\pi}{3}(2\alpha \hbar c/r_{0})^{3/2} m^{1/2}_{\rm red} \frac{dN_h}{dyd^2p_{T ,h}} \frac{dN_l}{dyd^2p_{T ,l}}
\label{coal}
\end{equation}

The enhanced atomic yield from Eq.~\eqref{coal} over that from Eq.~\eqref{ex:equation:baym} is:
\begin{equation}
\chi = \frac{1}{6\pi\zeta(3)}(\frac{2\hbar c}{m_{\rm red}r_{0}\alpha})^{3/2}.
\label{enh}
\end{equation}
For $r_{0}\simeq10$~fm, the enhancement factor for the $K$-$\mu$ atom is $\chi\simeq21.6$.

\begin{figure}[tbp]
\includegraphics[width=0.4\textwidth]{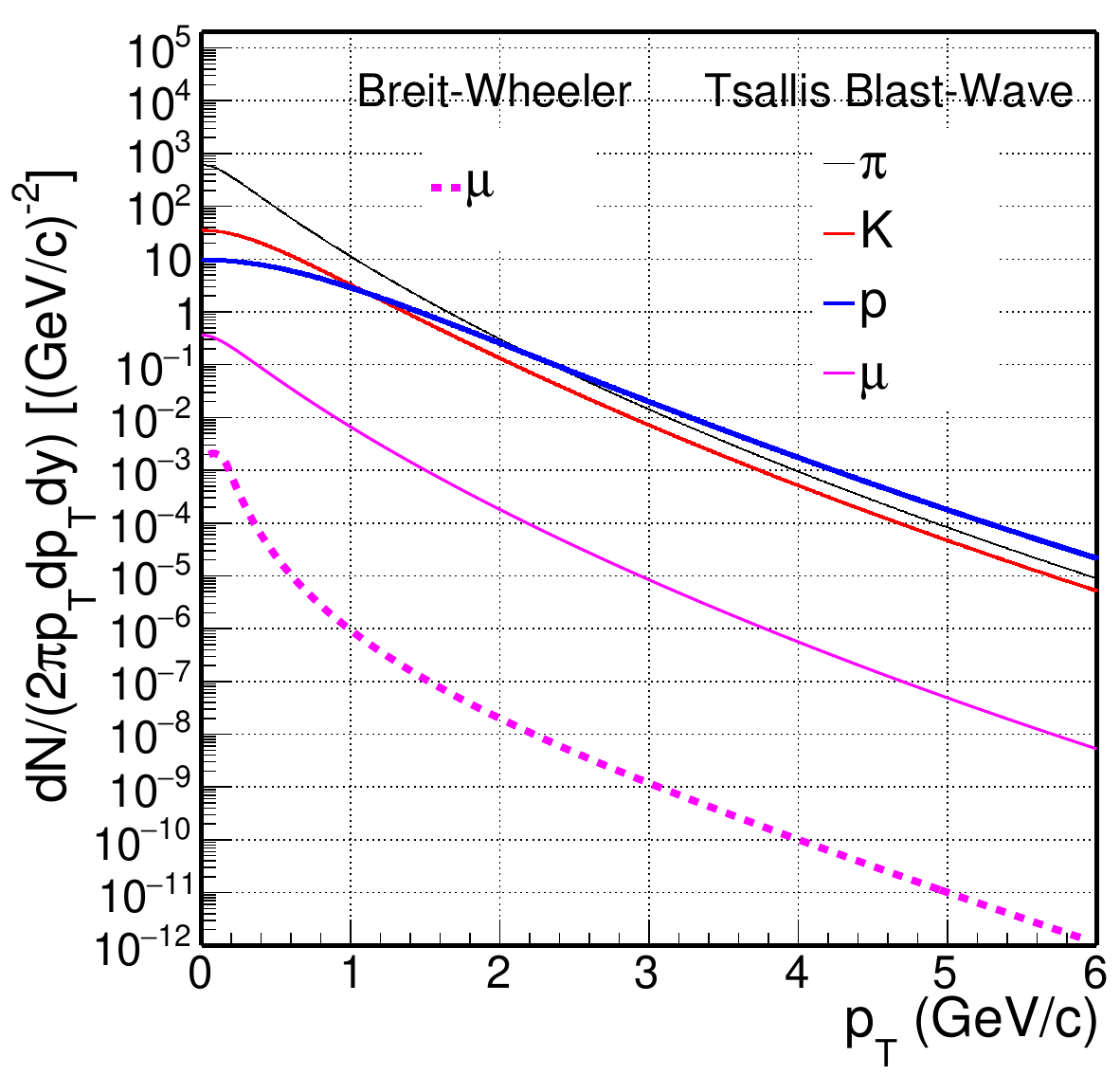}
\caption{Spectra of the hadrons as a function of transverse momentum in Au+Au collisions at 200 GeV with a centrality of 0-10\% from Tsallis Blast-Wave  model~\cite{Chen:2020zuw} and Breit-Wheeler process~\cite{Wang:2022ihj,Zha:2018tlq}. }
\label{ex:figures:hadron}
\end{figure}

\begin{figure}[tbp]
\includegraphics[width=0.4\textwidth]{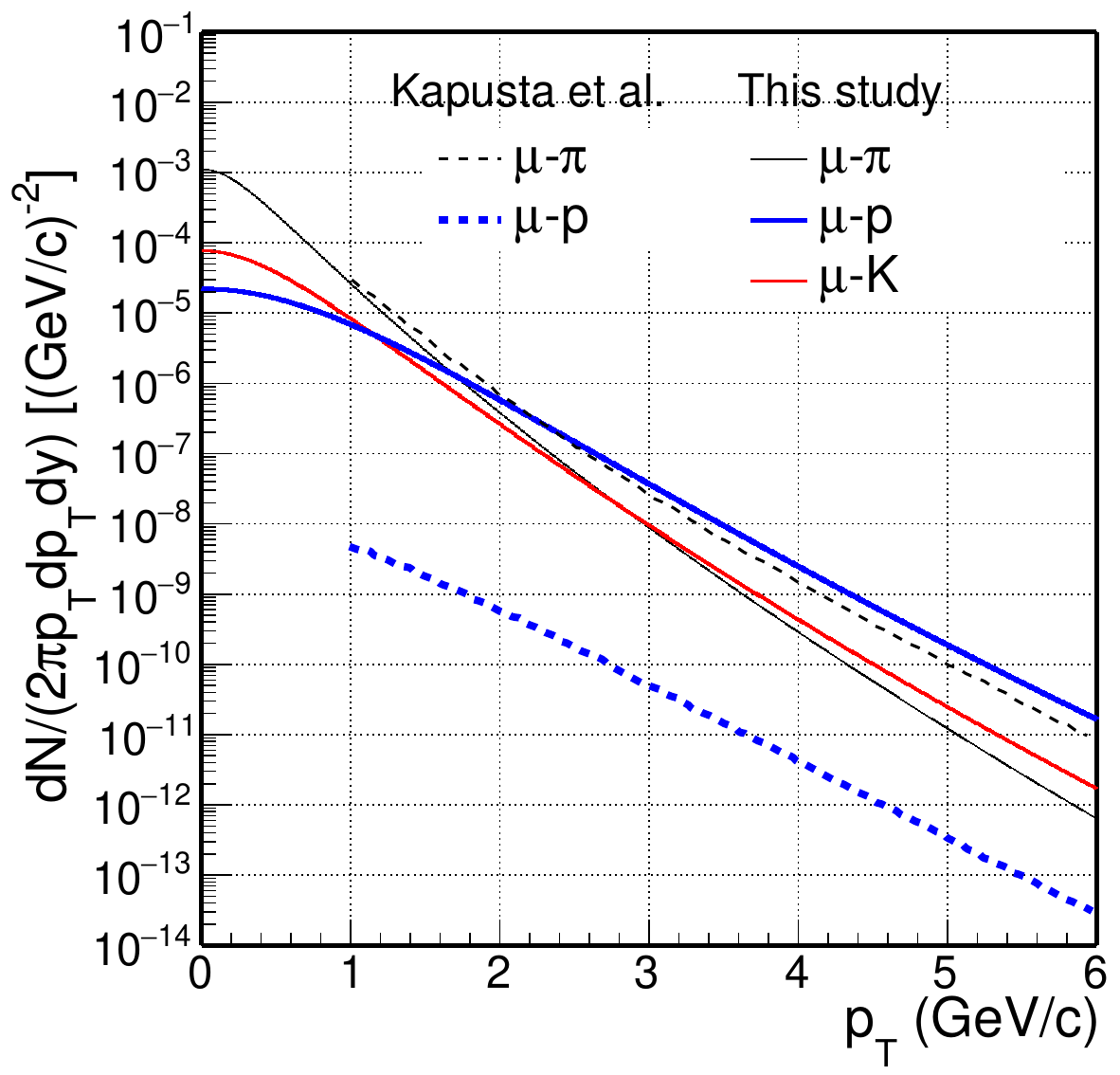}
\caption{Spectra of the muonic atoms as a function of transverse momentum in Au+Au collisions at 200 GeV with a centrality of 0-10\% from our estimate and from Kapusta {\it et al.}~\cite{Kapusta:1998fh}. }
\label{ex:figures:atom}
\end{figure}

\begin{figure} [htbp]
\includegraphics[width=0.4\textwidth]{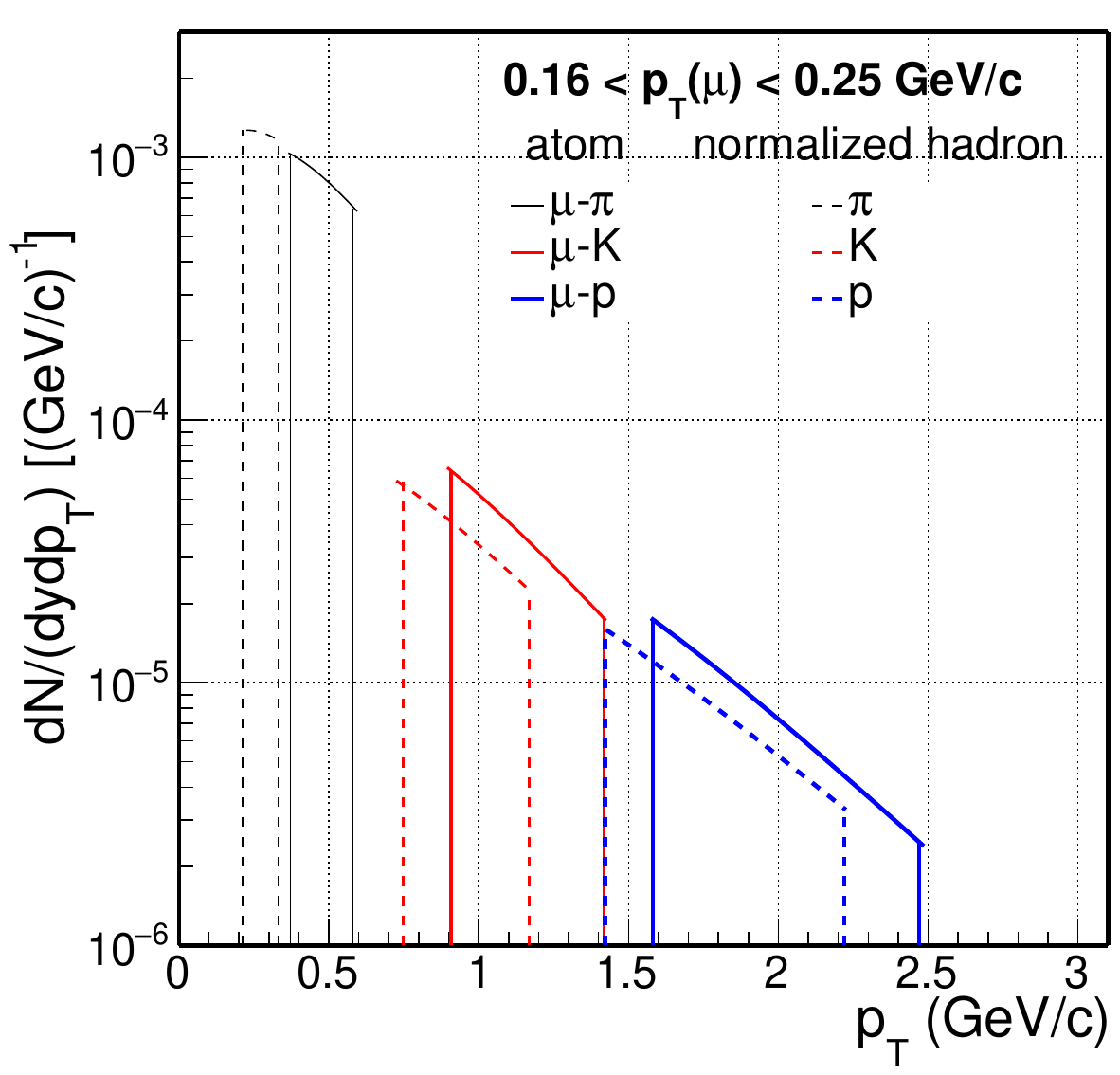}
\caption{Muonic atom and its normalized hadron daughter $p_T$ distributions with
muon momentum at $0.16<p_T<0.25$ GeV/$c$. }
\label{ex:figures:mupT}
\end{figure}

First, we use the Tsallis-Blast-Wave (TBW) model to estimate the yields of hadrons. The TBW model has effectively described the transverse momentum distribution of hadrons in heavy-ion collisions~\cite{Tang:2008ud,Shao:2009mu,Tang:2011xq,Ristea:2013ara}. The form of the TBW model is:
\begin{equation}\label{eq:TBW}
\begin{split}
   \frac{d^{2}N}{2\pi p_{T}dp_{T}dy}|_{y=0} & =A\int^{+Y}_{-Y}m_{T}\cosh(y)dy\int^{+\pi}_{-\pi}d\phi \\
  & \times\int^{R}_{0}rdr[1+\frac{q-1}{T}(m_{T}\cosh(y)\cosh(\rho) \\ 
  & -p_{T}\sinh(\rho)\cos(\phi))]^{-1/(q-1)},
\end{split}
\end{equation}
\noindent where $m_{\rm T} = \sqrt{m_{0}^{2} + p_{T}^{2}}$ is the transverse mass of particle with mass $m_{0}$ at a given transverse momentum $p_{T}$, $y$ is the source rapidity, and $\phi$ is the particle emission angle in the rest frame of thermal source. $q$ is the parameter characterizing the degree of non-equilibrium of the produced system, $T$ describes the temperature at which particles reach thermal equilibrium after a collision. $\rho = \tanh^{-1}(\beta_{s}(\frac{r}{R})^n)$ is the flow profile growing as the $n$th power from zero at the center of the collisions to $\beta_{s}$ at the hard-spherical edge ($R$, 1 is used in this study) along the transverse radial direction ($r$) and $\beta = \beta_{s}/(1 + 1/(n + 1))$ is the average flow velocity. We have used $n = 1$ in this study. 

In reference~\cite{Chen:2020zuw}, the TBW model has been applied to the midrapidity transverse-momentum spectra of identified particles measured at the RHIC. In this study, we selected Au+Au collisions at \sNN $=$ 200 GeV with a centrality of 0-10\%. The parameters $\beta$, $T$, and $q$ are 0.435, 0.118, and 1.036, respectively. Consequently, we obtained the transverse momentum distributions of $\pi$, $K$, and $p$, as shown by the black, red, and blue solid lines in Fig.~\ref{ex:figures:hadron}.

 To estimate the production rate of the thermal muon, we scale the pion spectrum in Fig.~\ref{ex:figures:hadron} down by a factor of ($(\alpha/\alpha_{s})^{2} \simeq 1/1690$) as a substitution for the unmeasured single muon spectrum, which is shown as magenta solid line in Fig.~\ref{ex:figures:hadron}. The production rate of muons from photo-induced processes is estimated using lowest-order QED~\cite{Wang:2022ihj,Zha:2018tlq}, which is shown as a magenta dashed line in Fig.~\ref{ex:figures:hadron}. As can be seen from the Fig.~\ref{ex:figures:hadron}, the contribution of muons from the Breit-Wheeler process is two orders of magnitude lower compared to the thermal contribution, and thus it has been neglected in this study.

Figure~\ref{ex:figures:atom} shows the invariant spectra of $\mu$-$\pi$, $\mu$-$K$, and $\mu$-$p$.
In addition to the enhancement factor $\chi$ in Eq.~\eqref{enh}, the $\mu$-$\pi$, \muk, and \mup\ have similar yields in our estimation, but the \mupi\ spectrum is significantly higher than \mup\ from the predictions of Kapusta and M\'ocsy~\cite{Kapusta:1998fh}.
This is mainly due to the different estimates of other hadron spectra and the low-$p_T$ muon yields they used. We emphasize that these estimates are intended to provide an order of magnitude for the feasibility study. This serves as guidance for the physics program, and is not intended to be a very vigorous theoretical prediction. The actual yields will be measured
experimentally and will provide first-hand single muon spectra.


Muonic atoms are created in the collisions as neutral particles.
They travel in a straight line from the collision vertex through the vacuum to the beam pipe wall, where the material of the
beam pipe dissociates the atoms into its muons and hadrons. In a solenoidal magnet field of a detector such as STAR, the oppositely charged
muon and hadron are bent in different directions entering the central tracker such as STAR's Time Projection Chamber (TPC)~\cite{Xin:2014lza,ex:star2010}. The signature of these atoms will be a V0 decay
vertex at the beam pipe with a muon and a hadron carrying opposite charges. To observe these atoms, we would need the
following:
\begin{itemize}
\item Dissociation of the atoms in detector material before the detector tracking and identification of the hadron and muon. It has been calculated by Prasad~\cite{Prasad:1979vn} that 0.01~inch of aluminum foil is sufficient to dissociate the atoms. This has been validated experimentally by Aronson {\it et al.} in 1982~\cite{Aronson:1982bz}. The aluminum foil is equivalent in terms of radiation length (0.3\%) to STAR's beryllium beam pipe.
\item Muon identification at the appropriate momentum. With the combination of ionization energy loss ($dE/dx$) measured by TPC and timing measurements by the Time-Of-Flight (TOF) detectors, STAR can cleanly identify muons at low momenta ($0.16<p_T<0.25$~GeV/$c$). This will allow for three different ranges of momenta of $0.4<p_T<0.6$~GeV/$c$, $0.9<p_T<1.4$~GeV/$c$ and $1.6<p_T<2.5$~GeV/$c$ for $\mu$-$\pi$, $\mu$-$K$, and \mup\ respectively. In these ranges, the pion, kaons and protons are comfortably within the particle identification range of the TOF detector. In addition, muons can be identified by the Muon Telescope Detector (MTD), albeit at higher momenta ($p_T>~1.5$ GeV/$c$). This means we will be able to identify \mupi\ at $p_T>3.5$~GeV/$c$. For \muk\ and $\mu$-$p$, the corresponding atom momentum is too high for its rate to be experimentally accessible.
\end{itemize}

Based on the muon identification range discussed in the above section, we can project the spectrum ranges for all three
species of atoms. Fig.~\ref{ex:figures:mupT} shows the corresponding atom and its daughter hadron $p_T$ ranges when
the muons are identified between $0.16<p_T<0.25$~GeV/$c$ with TPC and TOF detectors.


\hfill
\begin{table}[htbp]
\begin{ruledtabular}
\begin{tabular}{cccccc}
Atom & $\mu$ $p_T$ & Hadron $p_T$& Atom $p_T$& $dN_1/dy$ & $dN_2/dy$\\ \hline
\mupi & [0.16, 0.25] & [0.21, 0.33]& [0.37, 0.58] &$5\times10^{-6}$ & $2\times10^{-4}$\\
\muk & [0.16, 0.25]& [0.75, 1.17]& [0.91, 1.42]& $9\times10^{-7}$ &$2\times10^{-5}$\\
\mup & [0.16, 0.25]& [1.42, 2.22]& [1.58, 2.47]& $4\times10^{-7}$ & $7\times10^{-6}$\\
\mupi & $>1.5$ & $>2.0$ & $ >3.5$ & $3\times10^{-10}$ & $1\times10^{-8}$\\
\end{tabular}
\end{ruledtabular}
\caption{$p_T$ ranges for muonic atoms accessible to STAR and the integrated yield $dN/dy$ within the $p_T$ acceptances in central Au+Au collisions at 200 GeV. $dN_1/dy$ is the estimate from Eq.~\eqref{ex:equation:baym} and $dN_2/dy$ is the estimate from Eq.~\eqref{coal}.}
\label{ex:muonpt}
\end{table}

Table~\ref{ex:muonpt} lists the $p_T$ ranges for three types of atoms and their daughters identifiable by STAR and each with appreciable rates.
For example, the $\mu$-$\pi$ atoms production rate is about 200 candidates in one unit of rapidity in one million
central Au+Au collisions. The combined detection and identification efficiency is on the order of 10\% at low $p_T$~\cite{Wang:2023kzx}. Thus, a data sample of
500 million central events will have about 10000 such candidates. These measurements provide three different ways to measure the direct muon yields at low momentum using different muonic atoms.

In summary, RHIC and the STAR detector provide a unique environment for producing and detecting new particles. The measurements outlined in this paper not only provide potential discoveries of a new form of antimatter and exotic particles but also a unique tool for probing the properties of the quark gluon plasma.

\section*{Acknowledgments}
This work was supported in part by the Offices of NP within the U.S.\ DOE Office of Science under the contracts DE-FG02-89ER40531, DE-AC02-98CH10886 and DE-SC0005131, as well as the National Nature Science Foundation of China under Grant No.\ 12361141827.


\bibliography{ref}

\end{document}